\documentclass[twocolumn,showpacs,preprintnumbers,amsmath,amssymb]{revtex4}

\usepackage{subfigure}
\usepackage{graphicx}% Include figure files
\usepackage{dcolumn}% Align table columns on decimal point
\usepackage{bm}% bold math

\begin{document}

\title {Spin-dependent magnetoresistance and
spin-charge separation in multiwall carbon nanotubes}

\author{X. Hoffer, Ch. Klinke, J.-M. Bonard, L. Gravier}
  \affiliation{IPN, Facult\'e des
  Sciences de Base, EPFL, 1015 Lausanne, Switzerland.}

\author{J.- E. Wegrowe}
\email{wegrowe@hp1sesi.polytechnique.fr}
\affiliation{Laboratoire des Solides Irradi\'es, Ecole
polytechnique, 91128 Palaiseau Cedex, France.}

\date{\today}

\begin{abstract}
The spin-dependent transport in multiwall carbon nanotubes obtained by
chemical vapor deposition (CVD) in porous alumina membranes is
studied.  The zero bias anomaly is found to verify the predicted
Luttinger liquid power law.  The magnetoresistance at high fields
varies in sign and amplitude from one sample to the other, which is
probably due to the presence of dopant in the tube.  In contrast, the
magnetoresistance due to the spin polarized current is destroyed
in the nanotube as expected in case of spin-charge
separation.

\end{abstract}


\maketitle

Carbon multiwall nanotubes (MWNT) and single wall nanotubes (SWNT)
are considered as one of the most promising building blocks for
nanoelectronics and molecular electronics.  Among the large variety 
of possible applications, a
development in the framework of spintronics \cite{Prinz} is naturally
invoked, leading to studies of spin-dependent transport in carbon
nanotubes \cite{Alphenaar,Zhao1,Orgassa,Haruyama}.  In this
context, the spin dependent magnetoresistance
(SD-MR) of a MWNT contacted between two ferromagnetic electrodes has
been measured as a function of the magnetization direction of the
ferromagnetic contacts. Beyond the interest in spintronics
applications, the study of such magnetic systems allows the
investigation of fundamental questions about the role of the spin
degrees of freedom in quantum wires or Luttinger liquids (LL)
\cite{Voit,Lorenz}, where a specific behavior is expected due
to spin-charge separation \cite{Kane,Si,Balents,Mehrez}.

In the present work we observed LL-like behavior in
samples consisting of one or a few nanotubes connected to a tunneling
junction. We use the typical scaling law $G \propto V_{bias}^\alpha$
at high voltage bias $V_{bias}$, and $G \propto T^\alpha$ at low
voltage bias (G is the conductance, T the temperature and $\alpha$ is
the scaling coefficient), as discussed recently in the literature for
SWNT and MWNT \cite{Rque,Bockrath,Yao,Postma,Bachtold}.  This scaling
law originates in the framework of the Luttinger liquid (LL) theory
\cite{Kane,Kane1997,Egger} in 1D quantum wires, or intrinsic
Coulomb blockade in MWNT \cite{Egger2}, or in the framework of the
environmental Coulomb Blockade (ECB) theory applied to quantum wires
\cite{Bockrath2,Sonin2}. The spin-dependent magnetoresistance, 
related to the magnetic
hysteresis of the contacts, is measured for each sample and related
to the scaling coefficient $\alpha$ in order to observe an effect of the
expected spin-charge separation. Our measurements confirm previous
(negative) observations revealing that the SD-MR is very small
\cite{Orgassa,Haruyama} for all measured tubes; a significant signal 
is measured
only at vanishing bias \cite{Zhao1} or for small tubes where
the role of the junctions becomes significant. These measurements
indicate that the nanotubes destroy the spin polarization of
the current.

\begin{figure}
\centering
\includegraphics[width=0.45 \textwidth]{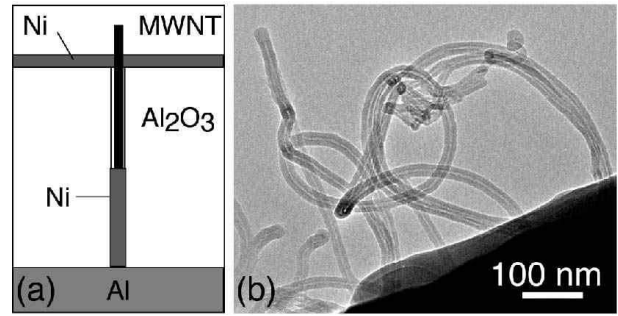}
\caption{\label{fig:epsart}  (a) Schematics of the Ni/MWNT heterostructure. (b) TEM
micrograph of NTs emerging from the pores at the membrane surface.}
\end{figure}

The MWNT were grown by a CVD technique in nanoporous alumina
membranes (see Fig.~1a)).  The membranes are obtained by anodization
of Al \cite{Haruyama}, and the length of the pores in the membrane is 1.5
$\mu $m.  Ni wires of controlled length are then electrodeposited in
the pores \cite{nous}.  The diameter of the MWNT are fixed at 23 nm $\pm$ 2
as shown by TEM (Fig.~1b).  The membrane is then exposed to 20 mbar acetylene
under 650$^\circ$C in a tube furnace during 5 min in order to activate
the catalytic growth of MWNTs from the top of the electrodeposited Ni
wires.  After nanotube deposition (see Fig.~1b), the tubes are kept in
air for a few minutes before a Ni layer (100 nm thickness) is
deposited with sputtering on top of the membrane.  The exposure to 
air leads to the
formation of a thin C-oxide layer which plays the role of a tunneling
junction at low temperature.  This tunneling junction is exploited for
tunneling spectroscopy.  We have studied a statistical ensemble (40)
of samples.  Each sample is defined by two sets of parameters, namely
the intrinsic parameters (length, purity of the tube, presence of
kinks) and the environmental conditions.  The latters are described in
terms of circuit theory by the impedance of tunneling junctions,
influence of other tubes contacted in parallel, and other sources of
dissipation.  The magnetic characteristics of the ferromagnetic
contacts also vary from one sample to the other, but the magnetic
configurations in such structures are well known from anisotropic
magnetoresistance (AMR) \cite{nous} and domain wall scattering (DWS)
\cite{nous} measurements.

The resistances range within 1k$\Omega $ to 100 k$\Omega $ for the 40
samples measured.  The contribution of both the tunneling junction and
the Ni wire to the resistance can be estimated from the R(T) profiles.
The scaling law is presented in Fig.~2 for a typical MWNT of length
of about 600 nm (sample {\bf A}, the resistance at 2K is 5 k$\Omega$).
The differential conductance $G=dI/dV$ is first plotted as a function
of the bias voltage V$_{bias}$ for different temperatures in
Fig.~2(a), showing a typical zero bias anomaly (ZBA).  The values at
zero bias G(V$_{bias}$=0) follow the power law $T^\alpha$ with
$\alpha$ = 0.23 (Fig.~2(b)).  In Fig.~2(c), $ GT^{-\alpha }$ is
plotted as a function of eV$_{bias}$/kT. All data collapse on a unique
curve, which indicates a LL-like behaviour.

\begin{figure}
\centering
\includegraphics[width=0.45 \textwidth]{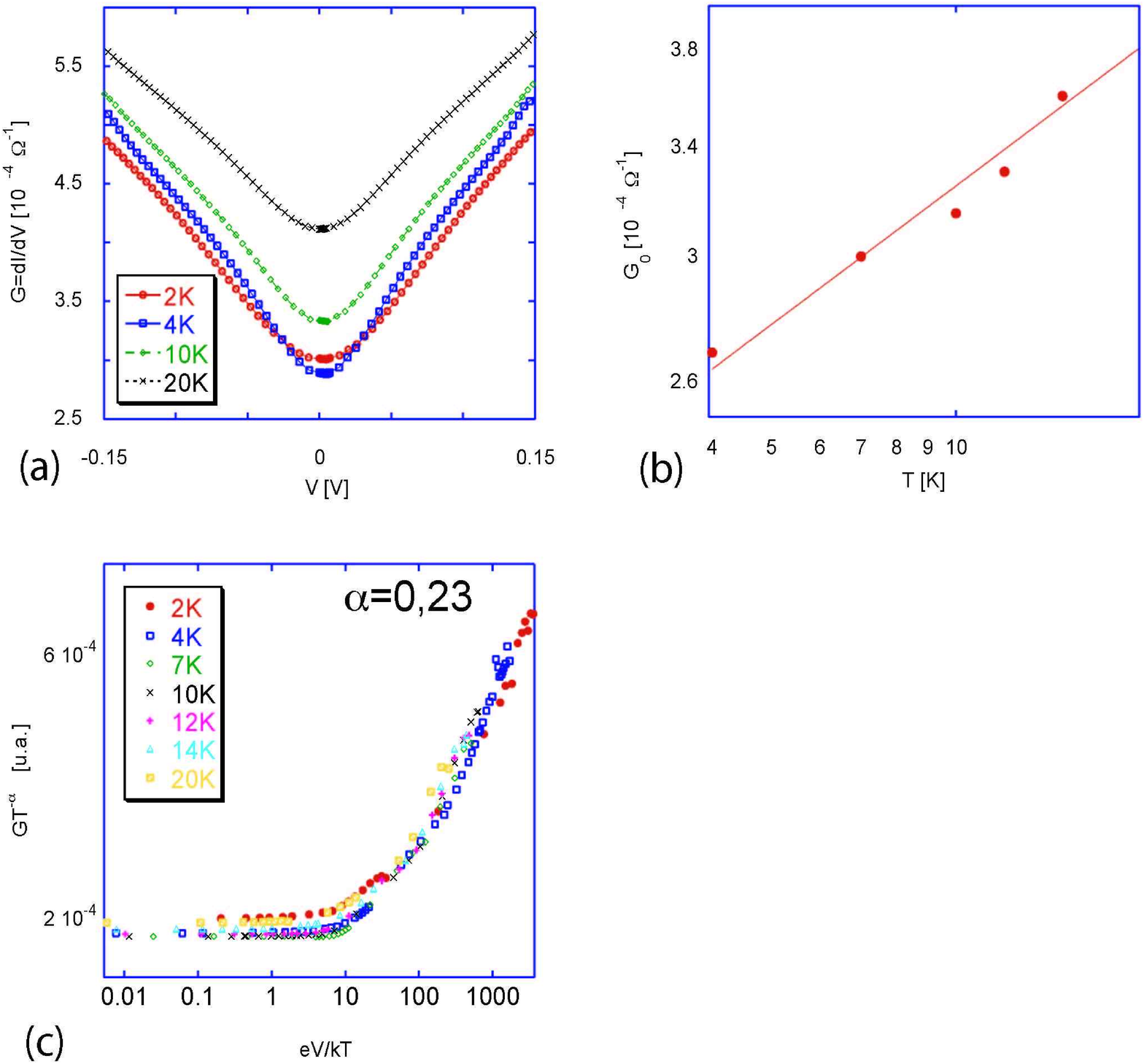}
\caption{\label{fig:epsart}  Sample A. (a) Conductance G (10$^{-4}$ $\Omega ^{-1}$) as 
a function of
$eV_{bias}$ for various temperatures.  (b) Log-Log plot of the temperature
dependence of the zero bias conductance as a function of the
temperature.  The line is a power law with $\alpha$=0.23.  (c) Scaling
GT$^{-\alpha}$ as a function of the ratio (eV$_{bias}$/kT).}
\end{figure}

Within the ensemble of 40 samples, 25 follow the scaling law for the
ZBA, which defines the coefficient $\alpha$ which are shown in Fig 3
(e.g. sample {\bf A},{\bf C} and {\bf D}).  Five samples show no
significant ZBA (Ohmic behavior : $\alpha$=0), and about 10 show a
strong ZBA, but without scaling law (e.g. sample {\bf B}).  The
coefficient $\alpha$ is distributed within the interval 0 $\leq \alpha
\leq$ 1 (Fig.  3).  In the framework of the transmission line approach
we have $ \alpha = 2 Re(Z)/R_{0} $, where $ R_{0}= h/e^2 $ is the
quantum resistance and Z is the impedance of the transmission line
(e.g. $ Z=R \approx \sqrt{L/C} $ with the impedance L and the
electrostatic capacitance C).  The theoretical value for a LL without
taking into account the environment is $ \alpha_{0} \approx 0.24 $
\cite{Kane,Kane1997,Egger,Bockrath2}.  In a first approximation, the
coefficient $\alpha$ is expected to be below the ideal value $
\alpha_{0}$ if the number of transmission modes N is enhanced
(typically $\alpha \propto \sqrt{1/N}$  \cite{Bockrath2}), e.g. due to
the large number of walls or impurities.  On the other hand, $\alpha$
should be larger than the bulk value if structural defects, like kinks, are
present \cite{Yao}.

The correlation between $\alpha$ and the length $ l $ of the tube is
plotted in Fig.~3. The length is estimated from the
deposition time of the Ni. Taking into account that the
probability of the presence of a kink or defect in the tube is 
proportional to the
length of the tube, we expect to measure scaling coefficients $\alpha
\le \alpha_{0}*$ for small $ l $ only, and $\alpha
\ge \alpha_{0}*$ for large $ l $. This is indeed observed in
Fig.~3, where there is no sample verifying $\alpha
\ge \alpha_{0}*$ below $l $=500nm (TEM measurements confirm that there
is a vanishing probability to find a kink at this scale), and no 
sample verifying $\alpha
\le \alpha_{0}*$ above $l $=900 nm.
The general tendency depicted in Fig.~3 is then
interpreted as
the presence of kinks or defects in long CVD grown nanotubes. The
large numbers of very small values of coefficient $\alpha$ for small
tubes (below 300 nm) can be
interpreted as an effect of screening of the tube by the contacts.

\begin{figure}
\centering
\includegraphics[width=0.45 \textwidth]{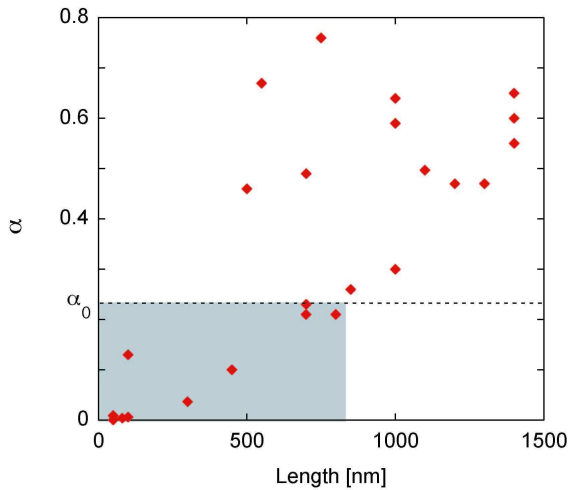}
\caption{\label{fig:epsart}   Scaling coefficient $\alpha$ as a  function of the length of
the MWNT. The grey zone corresponds to $\alpha
\le \alpha_{0}$, where $\alpha_{0}$ is the theoretical value calculated for a
Luttinger liquid.}
\end{figure}

If the transport properties at zero magnetic field can be described
with a single parameter $\alpha$, the MR properties are far more
versatile and various kinds of signals have been reported in the
literature \cite{Alphenaar,Zhao1,Orgassa,Haruyama}.  We separate
below two different types of MR, the direct MR due to the direct
action of the magnetic field on the charge, and the
spin-dependent magnetoresistance due to the
spin polarized current.  The MR has been measured for each sample with
a field perpendicular to the tube axis between $\pm$ 5T. At fields
above 1.5 T, the magnetization is saturated, and the profile gives the
direct MR of the nanotube.  As already reported in the literature,
the observed MRs (Fig.~4) are either positive (the resistance
increases with increasing magnetic field) or negative, varying from
one sample to the other.  Except for the sign, both positive and negative
MR are rather similar, with in some cases a transition from positive
to negative MR at low temperature (Fig.~4(a)).
Statistically, magnetoresistance at 2.5K is positive in 50 \% of the
samples and negative in 25 \%, while 25 \% have no measurable trend
(no MR at high field).  We do not observe
any correlation of the sign of the MR with the existence of the
scaling law, or with the value of the coefficient $\alpha$, or with
the temperature profile of the conductance.  Consequently, the origin
of this behavior must be ascribed to a small numbers of dopants, which
modify strongly the Fermi level from one sample to the other
\cite{Roche}, but do not modify significantly $\alpha$.

\begin{figure}
\centering
\includegraphics[width=0.45 \textwidth]{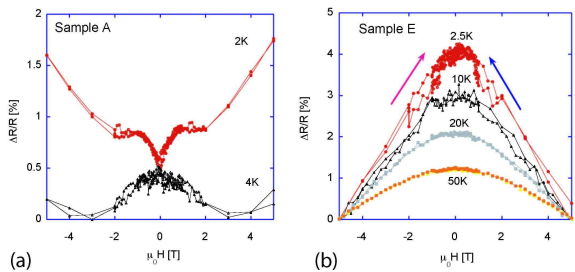}
\caption{\label{fig:epsart}  Magnetoresistance profile at large
fields for samples A (the curve is shifted by 0.5\% for clarity) and E.
  Sample A has a negative MR at low temperature and
sample E has a positive MR.}
\end{figure}

At low magnetic field, the magnetization of the wire (bottom
electrode) is oriented at an angle of 90$^\circ$ with respect to the Ni
layer (top electrode).  The magnetization of the Ni wire rotates
uniformly with increasing the applied field (following the
Stoner-Wohlfarth curve \cite{nous}), reaching the parallel
configuration at about 0.6 to 1.5 Tesla, depending on the length of
the Ni wire.  The measurements as a function of weak external
fields give then access to the conductance as a function of the angle
of the two ferromagnetic contacts (the so called SD-MR, i.e. the
effect of the spin polarization of the current).  Note that the
antiparallel configuration is not reached in these measurements.  The
efficiency of spin-injection process in a Ni nanowire (without carbon
nanotube) has been observed by a current-induced
magnetization-reversal effect in previous experiments \cite{nous}.
With respect to the SD-MR, the surprising result of the present study
is that the MR measured at reasonable current values (injected current
of about 1 $\mu$A) is always very small, below 1 \% of the total
resistance, whatever the length of the tube from 150 nm to 1500 nm 
(see also Ref.
\cite{Orgassa}).  Three different kinds of SD-MR are observed, which
are illustrated in Fig.~5 with samples {\bf A} to {\bf D}.
In sample {\bf B} (Fig.~5(b)), for small tubes the well-known AMR of 
the Ni wire is
measured where AMR signal is easily identified by
the typical shape as a function of the angle of the applied field
\cite{nous}.  The AMR is about 1.8 $\Omega$, i.e. 0.13 \% of the total
resistance, and $\alpha$=0.037.  The AMR signal shows that the current
is spin polarized and that the tube does not play any role in the MR.
In sample {\bf C} (Fig.5(c)), a second type of SD-MR hysteretic response
is also measured with $\Delta$R =20 $\Omega$ (about 0.7 \% of the total
resistance). In such cases, an important ZBA is observed,
but the conductivity cannot be scaled with power law.  The SD-MR is
observed at low temperature only, and disappears between 4K and 8K.
This behavior as a function of magnetic field and temperature is
typical for a very bad tunneling junction \cite{nous}.
In sample {\bf A} (presented in Fig.~2) and sample {\bf D} (for which
$\alpha$ = 0.59), a third type of SD-MR can be
measured {\it which depends on the direction of the current}.  Such
SD-MR is dramatically enhanced at very small or zero bias (see Ref.
\cite{Zhao1}) because it is due to the electrochemical potential
difference between the ferromagnet and the quantum wire.  This SD-MR
disappears at temperatures above 8K. This behavior can be understood
within the hypothesis of a non-equilibrium spin-injection that depends
on the incident spin-polarization of the current.  The spin-injection from
the ferromagnet, where the electrons are spin polarized, to the
quantum wire is not equivalent to the spin-injection from the quantum wire
(where there is no spin polarization of the current) to the
ferromagnet. These observations may
corroborate some predictions about non-equilibrium transport effects
due to spin-charge separation \cite{Balents}.

\begin{figure}
\centering
\includegraphics[width=0.45 \textwidth]{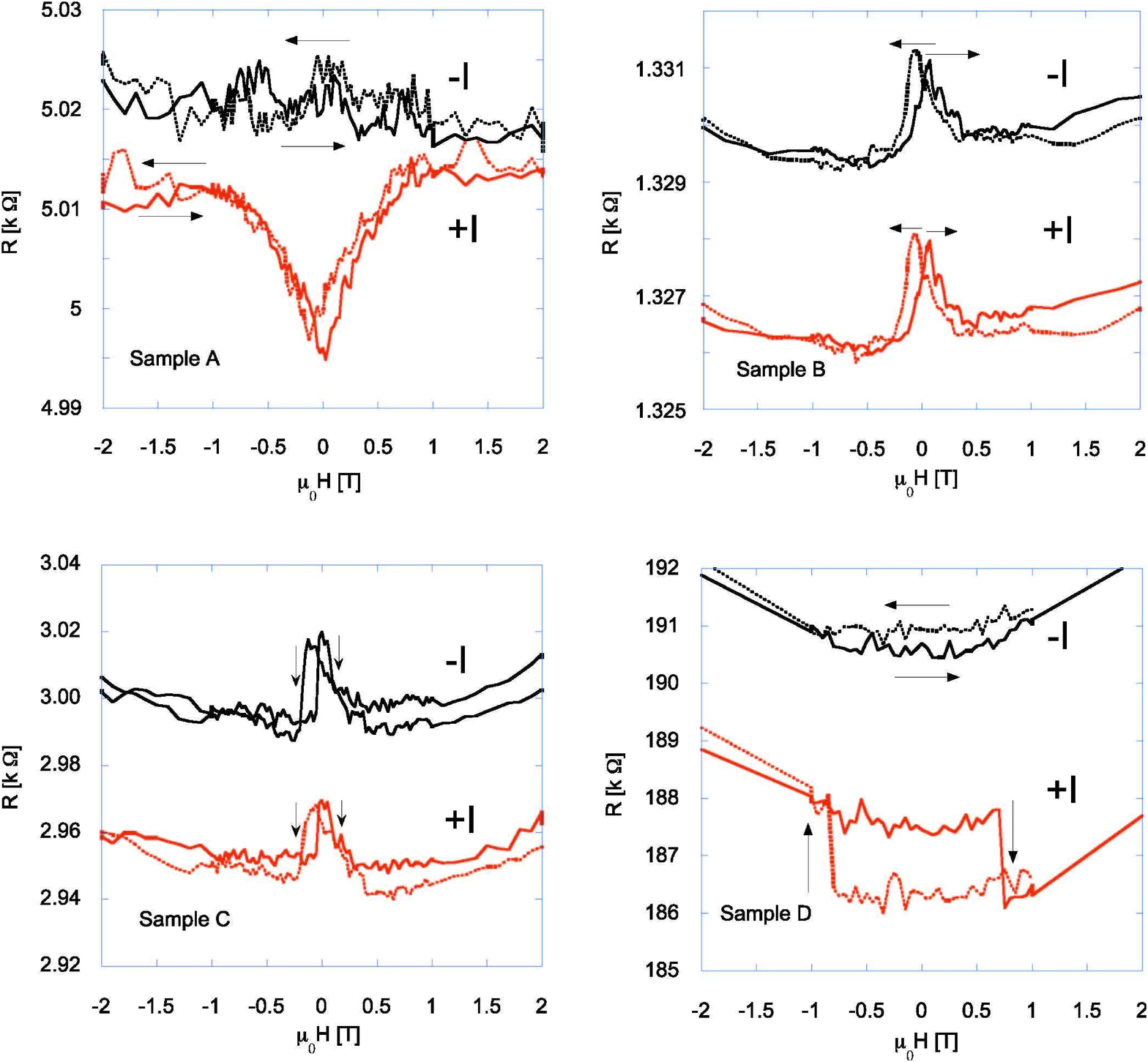}
\caption{\label{fig:epsart}   Spin-dependent magnetoresistance for four samples.(a) Sample
A, (b) Sample B : AMR response $\alpha$ = 0.037;
(c) sample C :  ZBA and no scaling property; (D) Sample D, $\alpha$ = 0.6.}
\end{figure}

In conclusion, we have measured the spin-dependent transport
properties of a series of MWNTs of various lengths, contacted by
ferromagnetic Ni electrodes.  Most of the samples exhibit a typical
scaling law behavior of the zero bias anomaly as a function of the
temperature.  We observed furthermore that the magnetoresistance due
to the spin polarization of the current is systematically destroyed in
the nanotube.  Only the AMR of the electrodes and some weak interface or
reservoir effects (spin-injection) are observed.  If we assume that
the observed scaling is due to
the manifestation of strong electron-electron interactions with
scattering between different modes, the destruction of spin-dependent
magnetoresistance leads then to a spin-diffusion
lengths below 150 nm.  This is in contradiction with the
semi-ballistic properties of nanotubes.  Consequently, these
results can be interpreted assuming that the scaling law $GT^{-
\alpha }(eV_{bias}/kT)$ originates from Luttinger Liquid behaviour,
and that the suppression of the spin dependent magnetoresistance is
due to spin-charge separation.

\acknowledgments

This work was supported by the grant n$^\circ$ 21-61550 of the Swiss
National Science Foundation. We are grateful to the Centre
Interd\'epartemental de Microscopie Electronique of EPFL (CIME-EPFL)
for access to electron microscopy facilities. We acknowledge
enlightening discussions with R. Egger, S. Roche, A. Bachtold and C.
Schoenenberger about spin dependent magnetoresistance in carbon
nanotubes.

%\end{multicols}

\end{document}